# Embedding the intrinsic relevance of vertices in network analysis: the case of centrality metrics

[1]Orazio Giustolisi*, [2]Luca Ridolfi, [3]Antonietta Simone

[1]Politecnico di Bari, via Orabona, 4, Bari, Italy, orazio.giustolisi@poliba.it
[2]Politecnico di Torino, Corso Duca degli Abruzzi, 24, Torino, Italy
[3]Politecnico di Bari, via Orabona, 4, Bari, Italy

**Abstract**

Complex network theory (CNT) is gaining a lot of attention in the scientific community, due to its capability to model and interpret an impressive number of natural and anthropic phenomena. One of the most active CNT field concerns the evaluation of the centrality of vertices and edges in the network. Several metrics have been proposed, but all of them share a topological point of view, namely centrality descends from the local or global connectivity structure of the network. However, vertices can exhibit their own intrinsic relevance independent from topology; e.g., vertices representing strategic locations (e.g., hospitals, water and energy sources, etc.) or institutional roles (e.g., presidents, agencies, etc.). In these cases, the connectivity network structure and vertex intrinsic relevance mutually concur to define the centrality of vertices and edges.

The purpose of this work is to embed the information about the intrinsic relevance of vertices into CNT tools to enhance the network analysis. We focus on the degree, closeness and betweenness metrics, being among the most used. Two examples, concerning a social (the historical Florence family's marriage network) and an infrastructure (a water supply system) network, demonstrate the effectiveness of the proposed relevance-embedding extension of the centrality metrics.

**Introduction**

Complex network theory (CNT) is today one the most active research fields allowing to analyze and predict the behavior of a wide variety of complex systems. CNT is becoming an emerging paradigm to study an impressive number of networked systems, ranging from physics to social sciences, form biology to infrastructure engineering. In this line, the growing computational power and data availability are giving a crucial contribution to the rapid growth of CNT tools for the network classification (Cohen and Havlin, 2010; Newman, 2018) community detection (Girvan and Newman, 2002; Fortunato, 2010), vertex and edge centrality assessment (Newman, 2018), vulnerability (Boccaletti et al., 2006), spatial and temporal evolution (Barthélemy, 2018), etc..

The evaluation of the centrality and the corresponding ranking have always been a key aspect in the network analysis. Freeman (1977) proposed the first comprehensive study of the─topological centrality, by identifying three main factors: (i) the degree as capability to spread information because of the number of local connections, (ii) the position in the network as importance of the vertices for the global exchange of information into the entire system, and (iii) the connections which identify preferential shortest paths into the network. Since then, centrality problem has attracted a lot of attention and many centrality metrics have been proposed (e.g., see Newman, 2018); examples are the Katz centrality (Katz, 1953), the betweenness centrality (Freeman, 1977), the closeness centrality (Freeman, 1977), the eigenvector centrality (Bonacich, 1987), the PageRank (Page et al., 1998), and the so-called Hub and Authorities (Kleinberg, 1999).

A fundamental characteristic of the efforts made to define and refine the centrality is that the concept of centrality has always been generally interpreted from a topological point of view, i.e. considering the position and connectivity of vertices and edges in the network, and not investigating the possible existence of an intrinsic vertex relevance independent from topology. This topology-based approach to the centrality problem is understandable as the identification of the centrality starting from the network connectivity structure is a considerable conceptual challenge. Nevertheless, the experience on networked systems shows that topology does not contain all the information, because the vertices can be characterized by a different intrinsic relevance that needs be considered. For example, in the social communities some people can act as influencer because of their activity in social media (i.e., influencer considering Instagram or Facebook), but others are influencers because of their reputation or representative institutional position (e.g., university rectors, presidents, prime ministers, etc.) in spite of their lower activity in social media. In the corresponding social networks, the first class of influencers is characterized by high degree of connections and their high topology-based centrality is correctly detected by classical CNT tools. The second class of influencers exhibits instead a lower degree of connections and standard CNT tools could not correctly quantify their actual centrality in the network. Infrastructure networks (e.g., water and energy distribution networks) are another example of the importance to consider the intrinsic relevance of vertices. In these cases, some strategic vertices, e.g., energy/water sources, hospitals, schools, administrative buildings, are characterized by a relevance independent from their topological position and connectivity; the intrinsic relevance needs to be embedded in the centrality evaluation to make useful CNT for such networked systems. We can state the identical intrinsic relevance of vertices is an inherent assumption CNT tools and that such assumption can hide a part of the network information, not allowing an effective analysis when vertices exhibit the heterogeneous intrinsic relevance.

The importance of relaxing the assumption of identical intrinsic relevance of vertices is demonstrated in this work focusing on three classical centrality metrics: the degree, the closeness, and the betweenness (Freeman, 1977; Newman, 2018). We here used the harmonic version of the closeness because it works for unconnected graphs (Marchiori and Latora, 2000).

We extend the standard definitions of these three metrics to embed the intrinsic relevance of vertices, showing that this information can increase the capability of centrality tools to assess the relevance of vertices and edges. The finding is that not considering the information about the intrinsic relevance of vertices can give misleading outcomes for some networked systems.

Actually, some researchers proposed to assign weights to vertices to analyse networks (Leung et al, 2011; Heitzig et al., 2012; Wiedermann et al., 2013; Topirceanu et al., 2018; Amano et al., 2018). Nevertheless, they focused on the assessment of vertices weight, dealing with the degree. Instead, Agryzkov et al. (2016) proposed a modified Adapted PageRank Algorithm and Agryzkov et al. (2019) modified the eigenvector centrality to establish a ranking of vertices in urban networks, embedding data about the presence of facilities (bars, coffees, restaurants, etc.) in the network topology. Both works, however, did not consider the metrics based on shortest paths and, therefore, the interaction between vertices. Differently, we here propose and discuss the importance of the vertices (their intrinsic relevance), by highlighting its different role with respect to the usual edges weight used to identify shortest paths. Then, we embed the intrinsic relevance of vertices in the shortest path-based centrality metrics, and in the degree centrality.

The work is organized as follows. The next section recalls the standard definitions of degree, (harmonic) closeness, and betweenness metrics and proposes the embedding of the intrinsic relevance of vertices in the standard formulations. Subsequently, the formulation of the intrinsic relevance is proposed and discussed. In the third section, the novel relevance-embedding centralities are applied to the historical Florence family's marriage network. Afterward, the novel extended betweenness is applied to an infrastructural (water supply) networked system. The last section reports the concluding remarks.

**Embedding vertex intrinsic relevance to degree, harmonic and betweenness centrality**

*Degree centrality*

We consider a network $N$ composed of $V$ vertices and $E$ edges. The degree $d(s)$ of a vertex $s$ is

$$d(s) = \sum_t A_{st} \quad t = 1...V \tag{1}$$

where $A_{st}$ is the adjacency matrix of the network. The standard degree of Eq. (1) assumes that all vertices are characterized by an identical intrinsic relevance. Such centrality informs on the vertex influence on its neighbours and, consequently, the metric exclusively depends on the (local) connectivity structure. We here relax the identical relevance assumption and undertake that each vertex has its own intrinsic relevance, $R_v$ ($v=1, …, V$). Furthermore, we introduce the function $f(R_s,$ $R_t$) that depends on the relevance of vertices $s$ and $t$. It weights the importance of the connection of the vertex $s$ with its neighbors. Hence, the definition of the degree can be extended to embed the information about the vertex intrinsic relevance as follows,

$$d(s) = \sum_t A_{st} f(R_s, R_t) \quad t = 1...V \tag{2}$$

To show the consequences of Eq. (2), we consider a pure regular network composed of four vertices and four edges as reported in Figure 1. We assume that the function $f$ is an increasing monotonic function and that $f(1, 1) = 1$ holds; in particular, in the present exercise we consider $f(R_s, R_t) = R_s \cdot R_t$.

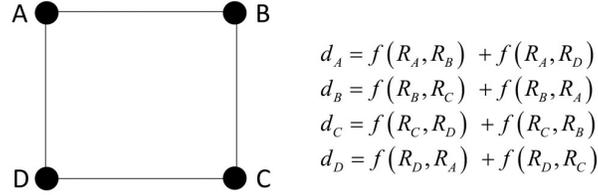

$$d_A = f(R_A, R_B) + f(R_A, R_D)$$
$$d_B = f(R_B, R_C) + f(R_B, R_A)$$
$$d_C = f(R_C, R_D) + f(R_C, R_B)$$
$$d_D = f(R_D, R_A) + f(R_D, R_C)$$

Figure 1. Regular network: the degree case.

In the standard condition ($R_A = R_B = R_C = R_D = 1$) Eq. (2) provides, as well as Eq. (1), the degree equal to two for each vertex of the regular network. We here double the relevance of A ($R_A = 2$) maintaining $R_B = R_C = R_D = 1$. The degree becomes: $d_A = 4$, $d_B = 3$, $d_C = 2$ and $d_D = 3$. Therefore, increasing the intrinsic relevance of the vertex A increases its relevance-embedding degree and that of the adjacent vertices B and D. It follows that embedding the intrinsic relevance in the degree allows to account for the vertex relevance together with the connectivity structure of the network, generating hubs. In the specific case, the regular network becomes more and more scale free.

*Harmonic centrality*

The harmonic version of the closeness $C(s)$ centrality of a generic vertex $s$ is

$$C(s) = \sum_t \frac{1}{dist_{s,t}} \quad t = 1...V \tag{3}$$

where $dist_{s,t}$ is the shortest distance between vertices $s$ and $t$. Also, the standard harmonic centrality assumes that all the vertices are characterized by an identical intrinsic relevance. Therefore, the values $C(s)$ exclusively depend on the connectivity structure of the network and, for weighted networks, on the edge weights driving the identification of the shortest paths and distance $dist_{s,t}$.

The harmonic centrality informs on vertex capacity to spread information through the network or to contact easily the other vertices of the network.

We embed the vertex intrinsic relevance for the harmonic version of the closeness centrality as,

$$C(s) = \sum_t \frac{f(R_s, R_t)}{dist_{s,t}} \qquad t = 1...V \tag{4}$$

To exemplify the impact of the vertex intrinsic relevance, we consider the same network and function $f(R_s, R_t)$ of the degree case. We also assume the edge weights, $w_{A-B} = 0.5$ and $w_{B-C} = w_{C-D} = w_{D-A} = 1$, to explain the different role of edge weights with respect to the vertex intrinsic relevance, as reported in Figure 2.

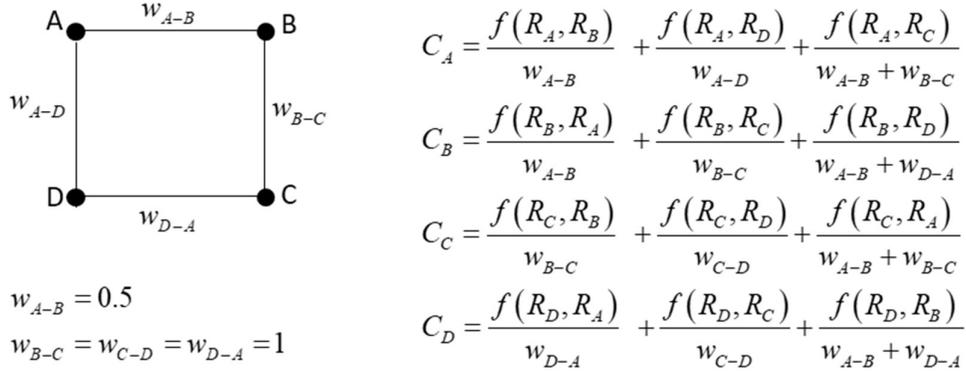

Figure 2. Regular network: the harmonic centrality case.

The standard assumption (i.e., $R_A = R_B = R_C = R_D = 1$) provides $C_A = C_B = 3.66$ and $C_C = C_D = 2.66$. In fact, the vertices A and B are characterized by a greater capacity to spread the information with respect to C and D because $w_{A-B}$ is lower than the other weights. Therefore, the connectivity structure of the network and the edge weights, determining the selection of the shortest path and the distance value between couples of vertices, characterize the vertex capacity to spread information. Notice that in the case of same edge weights, the harmonic centrality results $C_A = C_B = C_C = C_D$ because the network is regular also with respect to edge weights.

Similarly, to the degree case, we relax the assumption of identical relevance doubling $R_A (= 2)$ and maintaining unitary the others ($R_B = R_C = R_D = 1$). Eq. (4), made explicit in Figure 2 for the simple regular network, provides $C_A = 7.33$, $C_B = 5.66$, $C_C = 3.33$ and $C_D = 3.66$. Therefore, the intrinsic relevance of A increases its capacity to spread the information (i.e., the value of the harmonic centrality) and that of the nearest (as distance $d$) vertices.

This simple example shows how the intrinsic relevance concurs, together with the connectivity structure and edge weights (in the case of weighted networks), in increasing the capacity to spread information of the most relevant vertices and of the nearest ones.

*Betweenness centrality*

The standard–betweenness assigns the number of shortest paths traversing a vertex $v$ for all the couples of vertices $s$ and $t$ of the network. Mathematically, it reads

$$B(v) = \sum_{s \neq v \neq t \in N} \frac{\sigma_{s,t}(v)}{\sigma_{s,t}}, \tag{5}$$

where $B(v)$ is the betweenness of the vertex $v$, while $\sigma_{s,t}(v)$ and $\sigma_{s,t}$ are the number of shortest paths traversing the vertex $v$ and the total number of all shortest paths, respectively, for the each couple of vertices $s$ and $t$. Actually, more than one shortest path might exist for a couple of vertices $s$ and $t$; $\sigma_{s,t}$ accounts for this occurrence.

Similarly, the edge betweenness, $EB(e)$, is the sum of the fractions for all the couples of vertices $s$ and $t$ of the network traversing an edge $e$. The formulation is

$$EB(e) = \sum_{s \neq t \in N} \frac{\sigma_{s,t}(e)}{\sigma_{s,t}} \tag{6}$$

where $\sigma_{s,t}(e)$ is the number of shortest paths traversing the edge $e$ for all the couples of vertices $s$ and $t$ of the network.

Notice that, in the previous definitions (5) and (6), the fraction vanishes in the case of weighted networks. In fact, a single shortest path generally exists between each couple $s$ and $t$ in weighted networks; the vertex and edge betweenness formulations become, respectively,

$$\begin{aligned} B(v) &= \sum_{s \neq v \neq t \in N} \delta(v \in S_{s,t}) \\ EB(e) &= \sum_{s \neq t \in N} \delta(e \in S_{s,t}) \end{aligned} \tag{7}$$

where the Kronecker's $\delta$ function value is unitary if the vertices $v$ (or the edge $e$) is member of the shortest path $S_{s,t}$ and zero otherwise.

The betweenness informs in which elements of the network (vertices or edges) the information is more likely to pass.

We report in Figure 3 the way the standard betweenness is computed to illustrate the proposal of embedding the vertex intrinsic relevance. As shown in Figure 3(a), given the shortest path between two generic vertices $s$ and $t$, the betweenness of each traversed vertex (or edge, in the case of edge betweenness) has a unitary increment. After the identification of all the shortest paths between all the couples *s* and *t*, the betweenness assigns to each vertex (edge) the number of times has been traversed. Therefore, the values of the betweenness depend on the network connectivity structure and the presence of edge weights, which determine the shortest paths between couples of vertices.

We here propose to weight by $f(R_s, R_t)$ the shortest paths between *s* and *t*. I.e., we assume that its importance depends on the relevance of the two path-ending vertices *s* and *t*. Consequently, the betweenness of each element (vertex or edge) traversed by the path increases of a quantity $f(R_s, R_t)$. Figure 3(b) exemplifies the proposed assignment to internal elements of a shortest path.

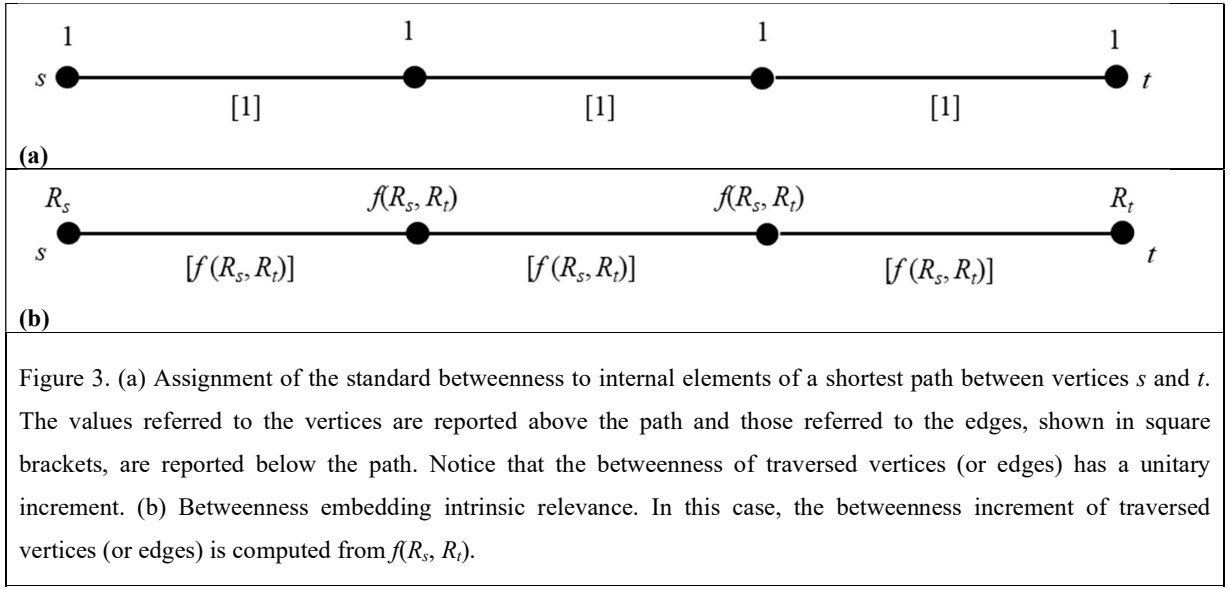

Figure 3. (a) Assignment of the standard betweenness to internal elements of a shortest path between vertices *s* and *t*. The values referred to the vertices are reported above the path and those referred to the edges, shown in square brackets, are reported below the path. Notice that the betweenness of traversed vertices (or edges) has a unitary increment. (b) Betweenness embedding intrinsic relevance. In this case, the betweenness increment of traversed vertices (or edges) is computed from $f(R_s, R_t)$.

Accordingly, the formulation of the betweenness embedding the intrinsic relevance becomes

$$B(v) = \sum_{s \neq v \neq t \in N} f(R_s, R_t) \delta(v \in S_{s,t})$$
$$EB(e) = \sum_{s \neq t \in N} f(R_s, R_t) \delta(e \in S_{s,t})$$
(8)

or, considering the most general case of Eqs. (5) and (6),

$$B(v) = \sum_{s \neq v \neq t \in N} f(R_s, R_t) \frac{\sigma_{s,t}(v)}{\sigma_{s,t}}$$
$$EB(e) = \sum_{s \neq t \in N} f(R_s, R_t) \frac{\sigma_{s,t}(e)}{\sigma_{s,t}}$$
(9)

To show the consequences of embedding the vertex intrinsic relevance, we use the same simple unweighted network of Figure 1. Figure 4 reports the shortest paths and the formulae to compute the betweenness and the edge betweenness. Notice that, for example, two paths exist from A and C and the factor ½ ($\sigma_{A,C} = 2$) accounts for that occurrence.

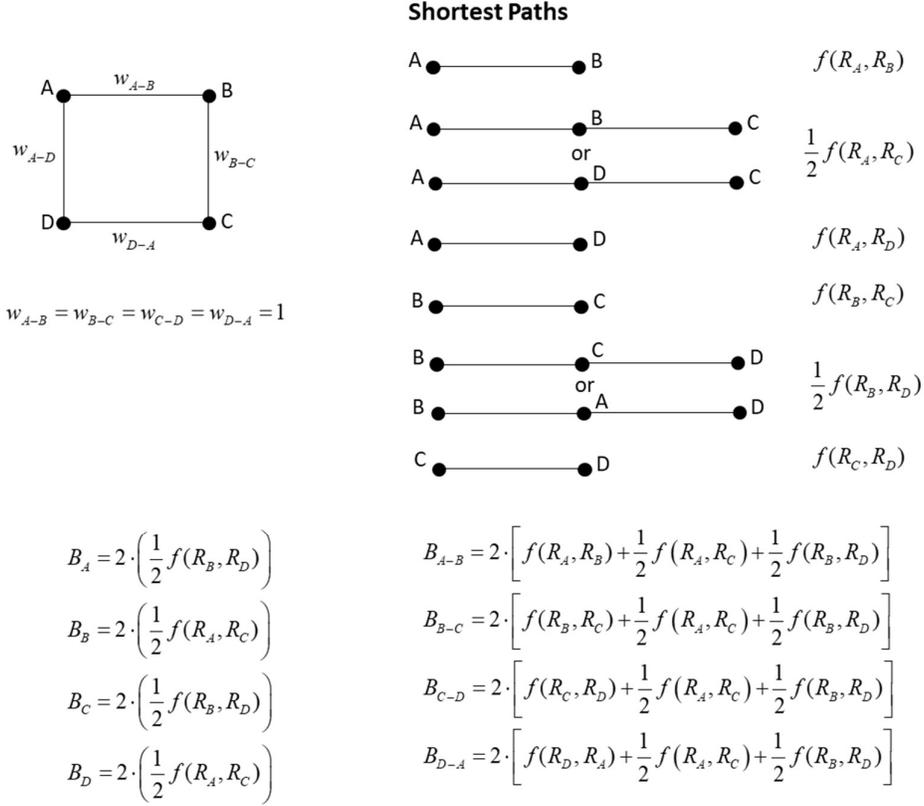

Figure 4. Regular network: the betweenness case using an unweighted network.

We perform the standard assumption $R_A = R_B = R_C = R_D = 1$, obtaining $B_A = B_B = B_C = B_D = 1$ and $EB_{A-B} = EB_{B-C} = EB_{C-D} = EB_{D-A} = 4$. Then, we double the relevance of A ($R_A = 2$) maintaining $R_B = R_C = R_D = 1$. The relevance-embedding betweenness becomes: $B_A = B_C = 1$ and $B_B = B_D = 2$, for vertices and $EB_{A-B} = EB_{D-A} = 7$ and $EB_{B-C} = EB_{C-D} = 5$ for edges. Consequently, embedding the intrinsic relevance increases the betweenness of vertices and edges which need to be traversed to reach the most relevant vertices.

Figure 5 reports the shortest paths and the formulae to compute the betweenness and the edge betweenness in the case of a weighted network, i.e. assuming the edge weights $w_{A-B} = 0.5$ and $w_{B-C} = w_{C-D} = w_{D-A} = 1$. Figure 5 is useful to show the different roles of edge weights and relevance of vertices in determining the betweenness.

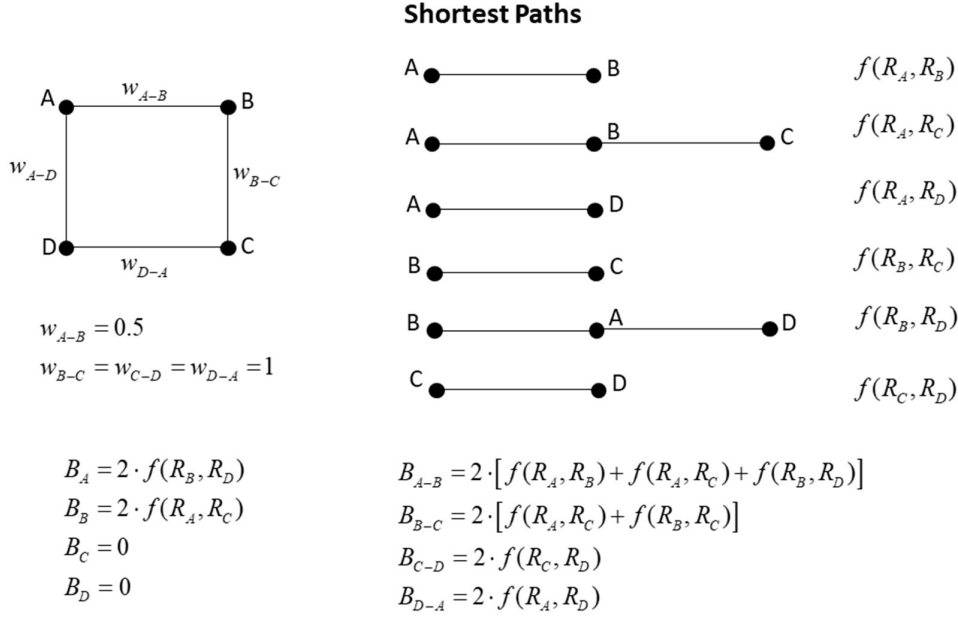

Figure 5. Regular network: the betweenness case using a weighted network.

The standard assumption ($R_A = R_B = R_C = R_D = 1$) provides betweenness values $B_A = B_B = 2$ and $B_C = B_D = 0$ for vertices and $EB_{A-B} = 6$, $EB_{B-C} = 4$ and $EB_{C-D} = EB_{D-A} = 2$ for edges. We here double the relevance of A ($R_A = 2$), the betweenness becomes: $B_A = 2$, $B_B = 4$ and $B_C = B_D = 0$ for vertices and $EB_{A-B} = 8$, $EB_{D-A} = 6$, $EB_{B-C} = 2$ and $EB_{C-D} = 4$ for edges. Again, embedding the intrinsic relevance increases the centrality of the vertices and edges which need to be traversed to reach the most relevant vertices.

The results for the simple regular network of Figures 4 and 5 confirm the findings of the degree and harmonic centrality: the vertex intrinsic relevance is an exogenous information to the standard network analysis, which can play an important role together with the network topology (connectivity structure and edge weights) in assessing the centrality ranking.

**Some comments about the vertex relevance and the function $f(R_s, R_t)$**

A first aspect that deserves to be underlined is that the vertex relevance is an exogenous information depending on the examined network and on the type of analysis to be performed. For example, in the case of infrastructure networks characterized by source vertices supplying water or energy to other vertices, the vertex relevance could be equal to the supplied water and energy for the source vertices and the demand for the other vertices. Moreover, the specific intrinsic relevance of some vertices (e.g., strategic buildings such as hospitals or schools) can be considered increasing suitably their corresponding relevance $R_v$ beyond the specific (water or energy) demand. Similarly, in the social networks the intrinsic importance of a vertex (e.g., an institutional vertex) can be embedded by introducing suitable values $R_v$ aiming to describe the "spectrum" of social intrinsic relevance of vertices. These are only few examples of relevance assignment; whose precise evaluation is problem-specific and then out of the scope of this work, maybe a further study of the CNT community once accepted the present proposal.

A second important point is the conceptual difference between the edge weights and the intrinsic relevance of vertices. The edge weights account for connections strength and play a role in determining the shortest path and the distance between vertices while the intrinsic relevance of vertices scales the inverse distance (harmonic centrality) or determines the importance of the shortest paths (betweenness).

About the function $f(R_s, R_t)$, we argue that its structure is problem-dependent, and that the selection of a specific function allows to embed a diverse exogenous information in the standard centrality metrics. For example, assuming $f(R_s, R_t) = R_s$ the degree $d(s)$ and closeness $C(s)$ are simply scaled by $R_s$ and the relevance of the vertex $s$ does not influence its neighbors, meaning that we are only increasing the degree and the capacity to transfer information of the specific vertex.

It follows that $f(R_s, R_t)$ can be inferred as a constitutive function defining the specific features of the network at assigned topology. In other words, our intrinsic relevance-based approach can be interpreted as the discrete analogue of a continuum model in physics, where one can describe several phenomena based on the same differential equation whereas the coefficients (here, the function $f(R_s, R_t)$) depend on the specific features of the problem at hands. Thus, the constitutive function $f(R_s, R_t)$ allows one to specify the physical networked system under analysis. The more detailed is the knowledge of the constitutive function the more precise will be the analysis. Furthermore, the availability of a theory about intrinsic relevance-based metrics allows one to design experiments to define the function $f(R_s, R_t)$ more suitable for specific physical networked systems.

However, two (already mentioned) conditions should be considered: (i) the relevance-embedding centrality metrics should be a generalization of the classic ones, therefore $f(1,1) = 1$ and (ii) $f$ should be monotonically increasing. The second condition follows from the fact that as the vertex relevance increase, it should increase the centrality metrics. Few examples of functions satisfying the previous constraints are: (i) $f(R_s, R_t) = R_s \cdot R_t$; (ii) $f(R_s, R_t) = (R_s + R_t)/2$; (iii) $f(R_s, R_t) = R_s$; (iv) $f(R_s, R_t) = \max[R_s, R_t]$; (v) $f(R_s, R_t) = \Sigma R_{path}$ and (vi) $f(R_s, R_t) = \Pi R_{path}$ (where $R_{path}$ indicates the relevance of vertices crossed in the path).

Finally, it is worth to notice that the values $f(R_s, R_t)$ can also be stored in a matrix having the same size of the adjacency one, having null diagonal elements. This paves a matrix-based definition of $f(R_s, R_t)$, generally asymmetric (i.e., $f(R_s, R_t) \neq f(R_t, R_s)$), built not considering an explicit function but observing at the specific analysed networked system.

In order to complete the assessment of the novel intrinsic relevance-based metrics proposed in this work, the supplementary material (see Supplementary Figures 1A- 24B) reports their application to two regular networks of 100 and 1,000 vertices and two random networks of the same size, generated using the Watts and Strogatz model (1998) with mean degree $d = 10$ and probability $p = \{0,1\}$. The standard betweenness and harmonic centrality were computed for each network, while the novel metrics were computed using each one of the six functions $f(R_s, R_t)$ previously defined. The intrinsic relevance was assigned to vertices adding to the basic unit value a random value sampled from a uniform distribution in the range $[0; d]$. At first, a fraction, $r$, of 10 % of vertices was randomly selected and the intrinsic relevance was randomly assigned. Then, the intrinsic relevance was randomly assigned to all the vertices ($r=100\%$) in the same range $[1; 1+d]$. Therefore, for each type of network (regular and random, with 100 and 1,000 vertices) we considered three cases: (i) all vertices share the same relevance equal to one, (ii) a fraction $r=10\%$ of vertices has random relevance in the range $[1; 1+d]$, and (iii) all vertices ($r=100\%$) exhibit random relevance in the range $[1; 1+d]$

The tests reported in the supplementary material (see Supplementary Figures 1A-24B) show a different finding for the relevance-embedding betweenness and harmonic centrality as consequence of the metrics different meaning.

About the betweenness centrality, the tests show that:
- for regular networks, the intrinsic relevance is always an important external information to characterize the specific features of the problem. I.e., the correlation between classic and extended metric is low;
- for random networks, the intrinsic relevance is an important external information when it concerns the 10% of vertices, while the role vanishes assigning a random relevance to all the vertices. I.e., the correlation between classic and extended metric is low for $r=10\%$ and increases for $r=100\%$;
- increasing the size of the random network, the role of the intrinsic relevance vanishes when randomly assigning it to all the vertices ($r=100\%$). I.e., the correlation between classic and extended metric increases;
- $f(R_s, R_t)$ based on the vertices of the path or based on the product of the ending vertices of the path emphasize the role of the intrinsic relevance;
- the intrinsic relevance is not correlated to the standard betweenness when $f(R_s, R_t)$ depends on the ending vertices of the paths. This is due to the specific meaning of the metric;

- the intrinsic relevance of vertices becomes to be correlated to the standard betweenness when $f(R_s, R_t)$ depends on all the vertices of the paths because the most relevant vertices contribute to the metric values when traversed.

About the harmonic centrality, the tests show that:
- the relevance-embedding metric is always an important external information to characterize the specific features of the problem, both for regular and random networks. I.e., the correlation between classic and extended metric is always low;
- the intrinsic relevance of vertices is quite always correlated to the harmonic centrality because the value of the metric for a vertex is at least scaled by its intrinsic relevance $f(R_s, R_t) = R_s$;
- the intrinsic relevance of vertices is not correlated to the harmonic centrality when $f(R_s, R_t)$ depends on all the vertices of the paths, because the intrinsic relevance of all the vertices in the shortest paths determines the metric value.

Finally, it is to bear in mind that the concept of intrinsic relevance for some networks refers to few vertices characterized by high values playing the role of "intrinsic relevance hub" as in the case of networked systems with sources (the following real cases will make clear this point).

**Two real examples: the Florence family's network and an infrastructure water supply system**

The effectiveness and importance of embedding the vertex intrinsic relevance is now shown by means of two different types of network. The first one is the Florence family's network (Padgett and Ansell, 1993); it is a social network frequently used as test case in the studies about centrality (e.g., Alvarez-Socorro et al., 2015; Latora et al., 2017; Sciarra et al., 2018). The second example is an infrastructure (water) network; this case will be useful to demonstrate the role played by source nodes and strategic customers.

In the following, we focus on the relevance-embedding betweenness centrality. The results corresponding to the relevance-embedding degree and harmonic centrality are reported in the Supplementary Material, although a discussion in the following text is delivered.

*Betweenness centrality for the Florence family's network*

The Florence family's network, reported in Figure 6, considers the seventeen most notables Renaissance Florentine families as vertices and the corresponding twenty-three marriages as edges. The network is not weighted. If the standard betweenness is computed (i.e., $R_v=1$ is assumed), we obtain the vertex and edge ranking shown in the figure 6 (see Supplementary Table S1 and S2 for the list of values), where the vertex centrality is represented by the family's emblem size and the edge betweenness is identified by the marriage link colour.

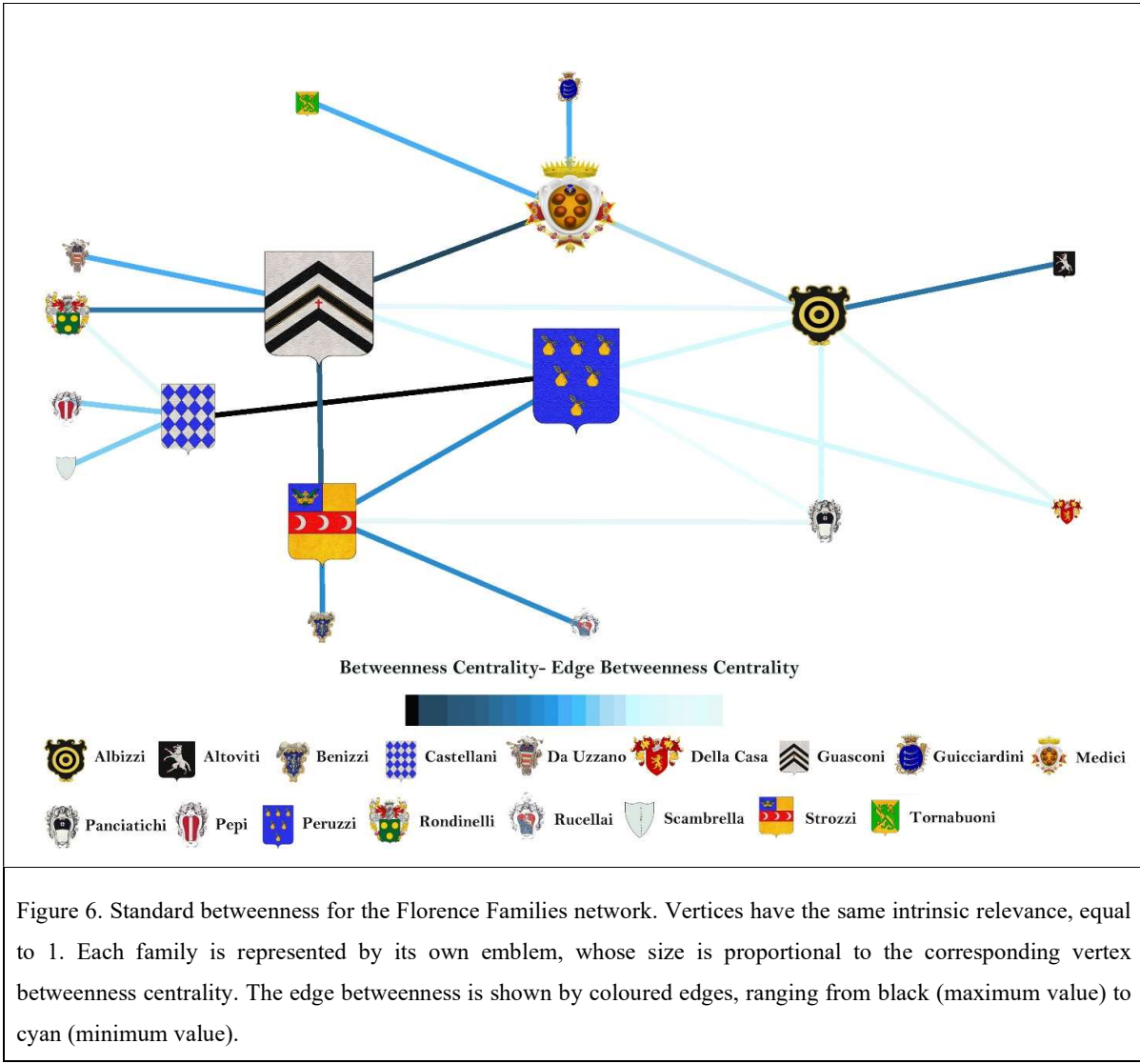

Figure 6. Standard betweenness for the Florence Families network. Vertices have the same intrinsic relevance, equal to 1. Each family is represented by its own emblem, whose size is proportional to the corresponding vertex betweenness centrality. The edge betweenness is shown by coloured edges, ranging from black (maximum value) to cyan (minimum value).

Figure 6 shows that Guasconi family is the most central family and the marriage between Peruzzi and Castellani families is the most important link. Nevertheless, this picture sounds quite strange; in fact, all history books have always reported that Medici is the most important Florentine family. Therefore, our CNT analysis based on the marriage topology is missing some key information. We think that this information is the intrinsic relevance of each family, evaluated by the gross wealth (i.e., the sum of marital, trade, partnership, bank and real estate relations) quantified in florins (Padgett and Ansell, 1993). The gross wealth of each family is reported close to the family name in Figure 7.

The novel betweenness embedding the family intrinsic relevance–is computed using the function $f(R_s, R_t) = R_s \cdot R_t$. Figure 7 shows that taking into account the vertex intrinsic relevance depicts a different (and more realistic) picture with respect to the standard betweenness: Medici becomes the most central family and the edge between Medici and Guasconi is the most important, i.e. the marriage between the two families (the list of values can be found as Supplementary Table S3 and S4) is the most central. The results in Figure 7 show that the family most likely to be traversed by the socio-economic activities is Medici, i.e. the most central in the socio-economic structure of the Florence families. The result is coherent with the History and it emerges considering the exogenous information of the intrinsic relevance in the standard CNT betweenness. In fact, although Medici has not the greatest gross wealth (i.e., intrinsic relevance), that family results the most central in the network because it is always traversed by the paths between the four of five richest Florentine families (Medici excluded), not being these four families linked by marriages. Namely, it is the coupling and interplay between topology and intrinsic relevance that make Medici the most central. Finally, the most important marriage (that one between Medici and Guasconi) explains and supports the centrality of Medici.

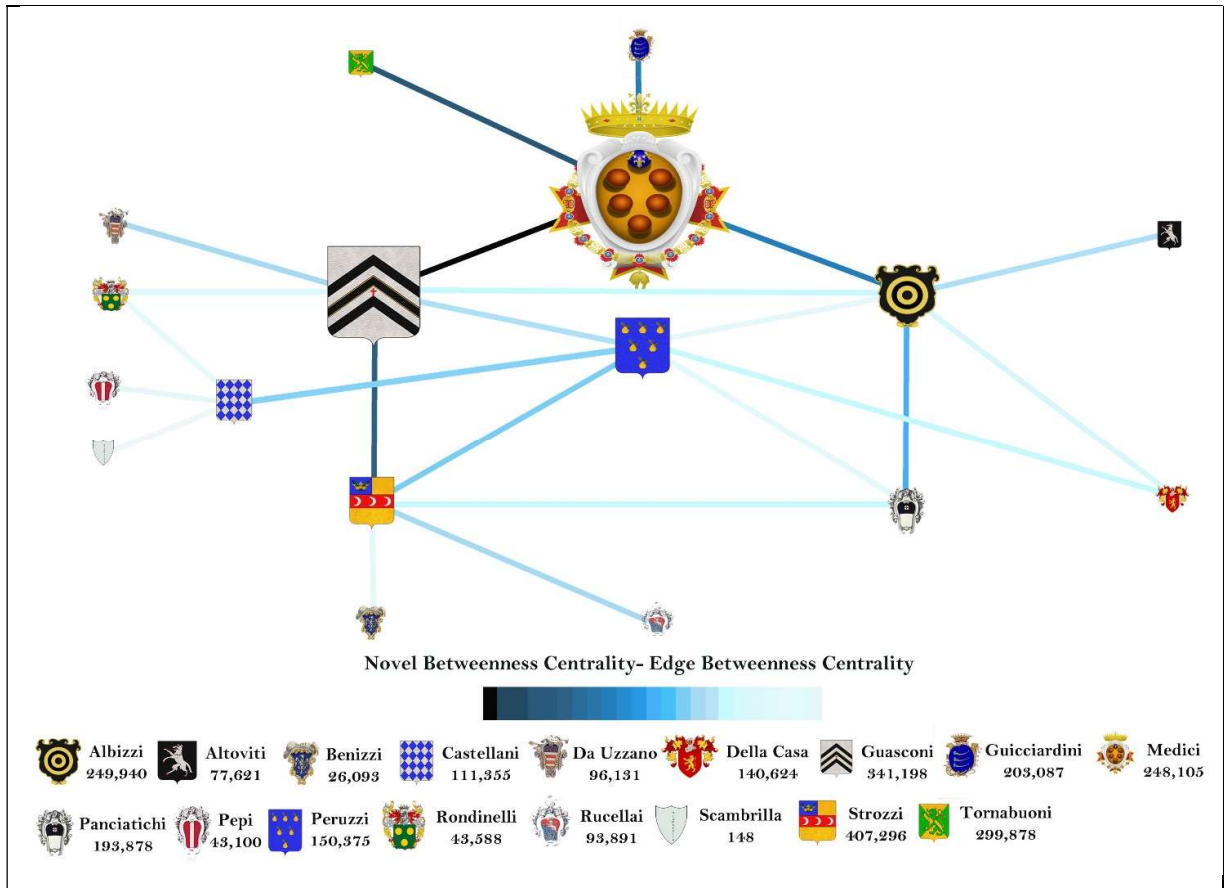

Figure 7. Proposed betweenness centrality, embedding the family intrinsic relevance of the Florence Family network. The intrinsic relevance corresponds to the family gross wealth (in florins) and its value is reported close to the name of each family. Emblem size and edge color (from black for maximum value to cyan for minimum value) highlight the vertex and edge centrality, respectively.

*Degree and harmonic centrality for the Florence Families network*

We here compute the standard degree and harmonic centrality and then the same metrics embedding the intrinsic relevance.

The degree ranges from 1 to 6 (see Supplementary Table S5 for the list of values) and the Guasconi, Albizzi and Peruzzi families are the most important hubs, being connected to six other families by marriages. The harmonic centrality ranges from 5.817 to 10.667 (see Supplementary Table S5 for the list of values). Guasconi and Peruzzi are the most central as capacity to spread socio-economic activities in the network closely followed by Peruzzi family.

The degree embedding intrinsic relevance ranges between 0 to 40.78E+11 (see Supplementary Table S6 for the list of values). Guasconi family becomes the most important socio-economic hub, being a rich family (intrinsic relevance equal to 341,198 florins) connected with six families by marriages. But the novel degree informs that Strozzi family, the richest one (intrinsic relevance equal to 407,296 florins) and connected with five families with marriages (including Guasconi), becomes the second socio-economic hub. Overall, the example shows that socio-economic influence of a family on its neighbours is relevant if the family is rich and well connected (by means of marriages) with other rich families.

The harmonic centrality ranges from 1,35E+08 to 6,08E+11 (see Supplementary Table S6 for the list of values). Guasconi remains the most central family to spread socio-economic activities through the network because of its economic relevance and topological position into the network of marriages. Again, Strozzi becomes the second in the ranking because of its economic relevance and position in the network of marriages.

Therefore, we can state that the degree and harmonic centrality embedding the intrinsic relevance of vertices allows to obtain a different information from the standard CNT analyses. In the present case study, the intrinsic relevance allows to integrate the economical information of the family relevance with the social information inherent the marriage network.

*The water distribution network*

The analysis of a distribution infrastructure highlights even more the need of embedding the intrinsic relevance of vertices for the centrality evaluation. To show this, here we refer to a simple water supply network composed of twenty-four vertices (that represent twenty-three demand nodes and one source node) and thirty-four edges corresponding to the pipes for the water distribution (see Figure 8) (Giustolisi and Ridolfi, 2014). The edges are weighted with the length of pipes to account for the fact that it is a spatial network. We here focus on the edge centrality because the pipes are the relevant components for distribution systems (Giustolisi et al., 2019).

If the standard edge betweenness is computed – i.e., the same intrinsic relevance is assigned both to source and demand nodes – the outcomes are those shown in the Figure 8 (left): the most important edge is the number 4 followed by the number 17 (the list of values is reported in the Supplementary Table S7 for the list of values). However, these results are unrealistic; in fact, it is evident that the most important edge for the distribution network corresponds to the pipe connecting the network to the source of water (the edge 34 connected to the vertex 24) because all the paths from the source of water always traverse that edge. Once again, we are missing information during the CNT analysis, i.e. the intrinsic relevance of the source vertex (node 24), which supplies water to the twenty-three demand nodes. In order to make centrality assessment more realistic, the intrinsic relevance is assigned equal to: (i) the water demand for each vertex corresponding to a demand node and (ii) the sum of supplied water for the source node. Figure 8 (right) reports the edge ranking as resulting from the edge betweenness embedding the vertex intrinsic relevance (we assume $f(R_s, R_t) = R_s \cdot R_t$). The demands, in [l/s], are reported in parenthesis close to each vertex. Differently from the standard betweenness, the novel edge betweenness correctly identifies the pipe 34 as the most important edge, consistently with the hydraulic engineering knowledge (see the Supplementary Table S8 for the list of values).

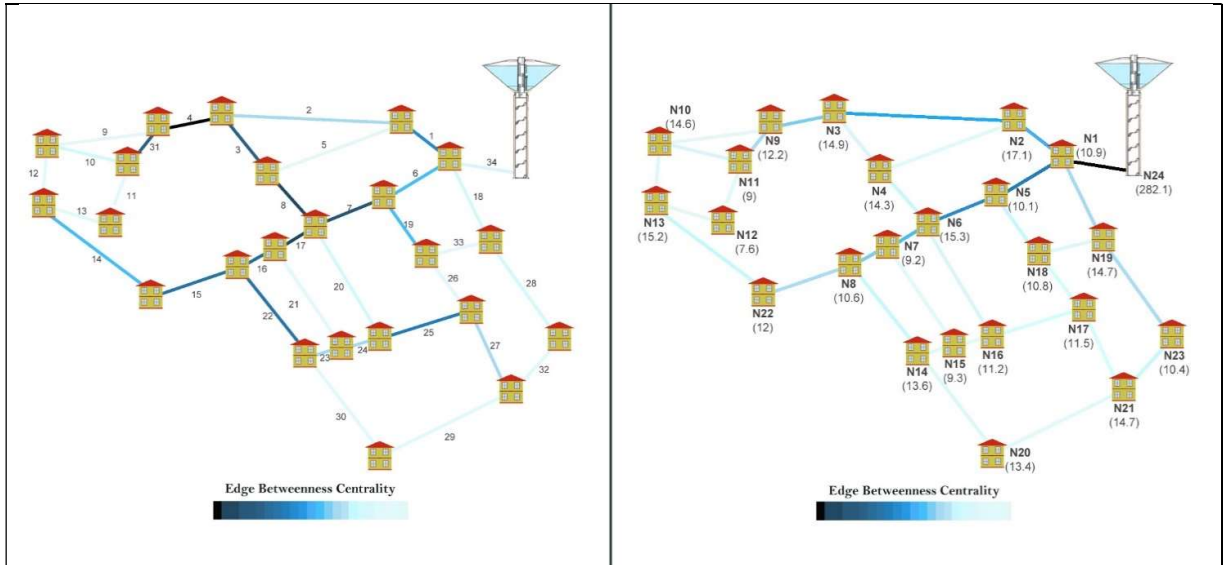

Figure 8. (left) Standard edge betweenness for a water supply network. Source and demand nodes have the same intrinsic relevance equal to 1. The edge betweenness is identified by coloured edges ranging from black (maximum value) to cyan (minimum value). The most important pipe is the number 4. (right) Proposed edge betweenness embedding the intrinsic relevance in terms of demand for each node and as sum of demands for the source node (values in l/s are reported close to each node). The pipe 34 results the most important edge.

We here consider the same water supply network with the further assumption that a strategic consumer, a hospital at the demand node 10, is present. Aiming to consider the peculiarity and importance of the hospital (whose failure to supply water would cause serious damage), the intrinsic relevance is augmented to ten times of its demand. Figure 9 reports the novel edge betweenness embedding the intrinsic relevance of the hospital (see the Supplementary Table S9 for the list of values): the edge 34 results again the most important pipe, but the main path is now identified by vertices 24-13 and not 24-8, as it was instead obtained when intrinsic relevance of vertex 10 was equal to its demand (see Figure 8, right panel). This fact demonstrates that the function $f(R_s, R_t) = R_s \cdot R_t$ correctly emphasizes the most important pipes for water supplying security at the hospital.

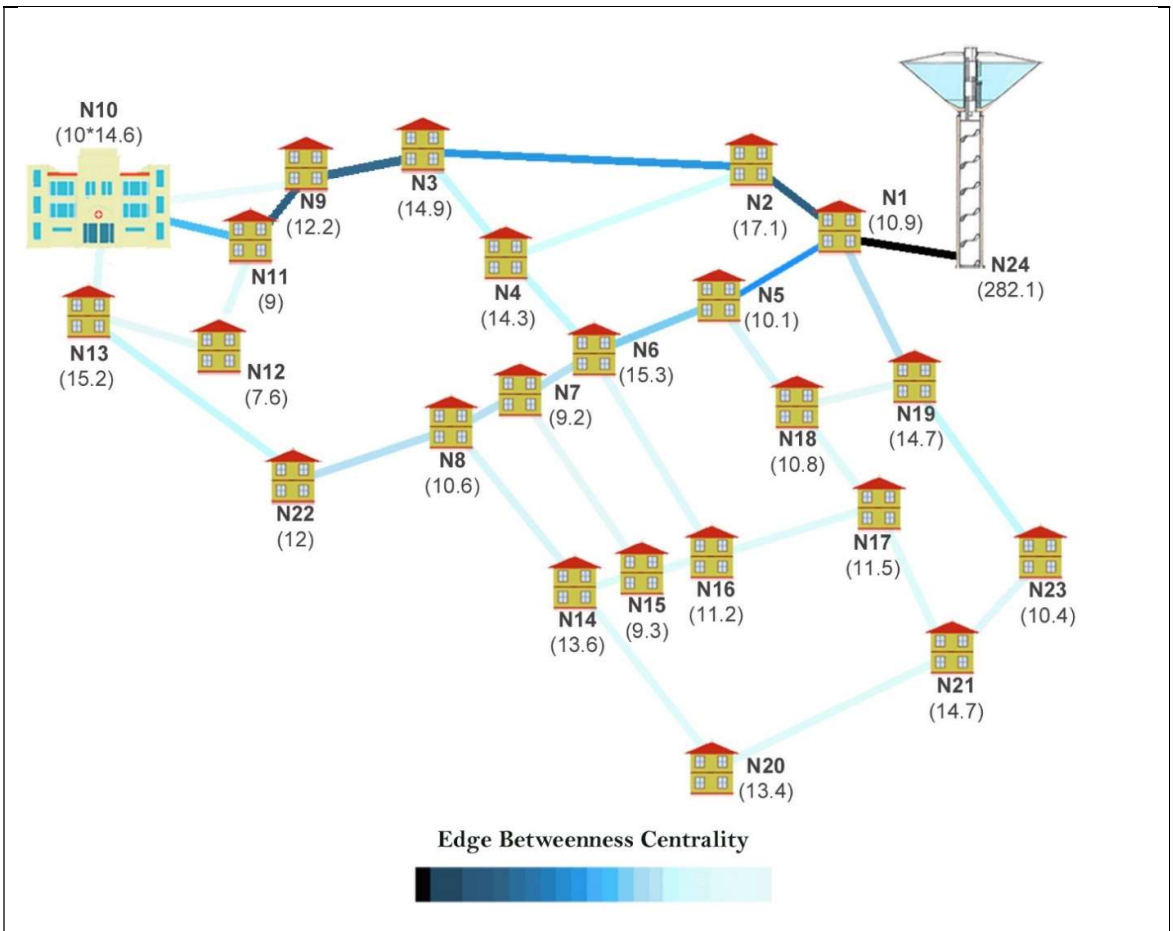

Figure 9. Intrinsic relevance-embedding edge betweenness when a hospital is present in the vertex 10. Its intrinsic relevance is ten times its demand. The proposed edge betweenness again identifies pipe 34 as the most important, but the main path in the network changes in favor of those connecting the source node and the hospital.

**Conclusions**

The aim of the present work is to draw attention to the importance of embedding the information about the intrinsic relevance of vertices into CNT tools to enhance the network analysis. The usual assumption of identical relevance of vertices makes the centrality assessment only founded on topological features (on network connectivity structure and, possibly, on edge weights) causing the ineffectiveness of the network analysis when such exogenous information is relevant.

The information about intrinsic relevance can refer to different occurrence: e.g., institutional roles, strategic values of the vertices (e.g., hospitals, embassies, schools, stations, etc. in the infrastructure networks), functions of the vertices in the network (source vs. demand nodes), etc.. The key idea of the proposal of embedding intrinsic relevance in CNT is to avoid losing exogenous information about vertices.

We focused on the degree, (harmonic) closeness and betweenness metrics, being among the most used, proposing relevance-embedding formulations and demonstrating their effectiveness. In fact, using two examples concerning a social and an infrastructure network, we demonstrated the crucial role played by intrinsic relevance in assessing the correct vertex and edge ranking.

We think that capturing the interplay between network topological structure and intrinsic relevance of vertices can open novel interesting paths in the CNT studies and applications to real problems. Our work wishes to be a step in this direction.

**Author contributions**

O.G. proposed the idea, O.G., L.R. and A.S. developed the analytical theory of the idea and conceived the case studies, A.S. designed the case studies, all authors contributed to the writing of the article.

**Additional information**